\documentclass[twocolumn,showpacs,superscriptaddress,amsmath,amssymb]{revtex4}
\usepackage{graphicx}
\usepackage{dcolumn}
\usepackage{bm}
\usepackage{color}

\def\bd{
\begin{document}} \def\ed{\end{document}}
\def\bmp{\begin{minipage}} \def\emp{\end{minipage}}
\def\bcc{\begin{center}} \def\ecc{\end{center}}     \def\npg{\newpage}
\def\beq{\begin{equation}} \def\eeq{\end{equation}} \def\hph{\hphantom}
\def\be{\begin{equation}} \def\ee{\end{equation}} \def\r#1{$^{[#1]}$}
\def\n{\noindent} \def\ni{\noindent} \def\pa{\parindent}
\def\hs{\hskip} \def\vs{\vskip} \def\hf{\hfill} \def\ej{\vfill\eject}
\def\cl{\centerline} \def\ob{\obeylines}  \def\ls{\leftskip}
\def\underbar#1{$\setbox0=\hbox{#1} \dp0=1.5pt \mathsurround=0pt
\underline{\box0}$}   \def\ub{\underbar}    \def\ul{\underline}
\def\f{\left} \def\g{\right} \def\e{{\rm e}} \def\o{\over} \def\d{{\rm d}}
\def\vf{\varphi} \def\pl{\partial} \def\cov{{\rm cov}} \def\ch{{\rm ch}}
\def\la{\langle} \def\ra{\rangle} \def\EE{e$^+$e$^-$} \def\pt{p_{\rm T}}
\def\pti{p_{{\rm T},i}} \def\yti{y_{{\rm T},i}}
\def\ptj{p_{{\rm T},j}}\def\mt{m_{\rm T}} \def\yt{y_{\rm T}} \def\vt{v_{\rm T}}

\def\bitz{\begin{itemize}} \def\eitz{\end{itemize}}
\def\btbl{\begin{tabular}} \def\etbl{\end{tabular}}
\def\btbb{\begin{tabbing}} \def\etbb{\end{tabbing}}
\def\beqar{\begin{eqnarray}} \def\eeqar{\end{eqnarray}}
\def\\{\hfill\break} \def\dit{\item{-}} \def\i{\item}
\def\bbb{\begin{thebibliography}{9} \itemsep=-1mm}
\def\ebb{\end{thebibliography}} \def\bb{\bibitem}
\def\bpic{\begin{picture}(260,240)} \def\epic{\end{picture}}
\def\akgt{\cl{\bf ACKNOWLEDGMENTS}}
\def\fgn{\noindent{\bf\large\bf figure captions}}
\def\m1{\langle N_p\rangle} \def\u2{\langle N_{\bar p}\rangle} \def\Nap{N_{\bar
p}}
\def\lan{\langle}
\def\ran{\rangle}
\def\p{\pi}
\def\ifmath#1{\relax\ifmmode #1\else $#1$\fi}%
\def\rc{\ifmath{{\mathrm{c}}}}
\def\cut{\ifmath{{\mathrm{cut}}}}
\def\rF{\ifmath{{\mathrm{F}}}}
\def\rK{\ifmath{{\mathrm{K}}}}
\def\rp{\ifmath{{\mathrm{p}}}}
\def\rt{\ifmath{{\mathrm{t}}}}
\def\LAB{\ifmath{{\mathrm{LAB}}}}
\def\cut{\ifmath{{\mathrm{cut}}}}
\def\beq{\begin{equation}}
\def\eeq{\end{equation}}

\newcommand{\cinst}[2]{$^{\mathrm{#1}}$~#2\par}
\newcommand{\crefi}[1]{$^{\mathrm{#1}}$}
\newcommand{\crefii}[2]{$^{\mathrm{#1,#2}}$}
\newcommand{\crefiii}[3]{$^{\mathrm{#1,#2,#3}}$}
\newcommand{\HRule}{\rule{0.5\linewidth}{0.5mm}}

\bd
\title{Impact of Limited Statistics on the Measured Hyper-Order Cumulants of Net-Proton Distributions in Heavy-Ion Collisions}

\author{Lizhu Chen}\email{chenlz@nuist.edu.cn}
\affiliation{School of Physics and Optoelectronic Engineering, Nanjing University of Information Science and Technology, Nanjing 210044, China}
\affiliation{Key Laboratory of Quark and Lepton Physics (MOE) and
Institute of Particle Physics, Central China Normal University, Wuhan 430079, China}
\author {Ye-Yin Zhao}
\affiliation{School of Physics and Electronic Engineering, Sichuan University of Science and Engineering (SUSE), Zigong 643000, China}
\author{Yunshan Cheng}
\email{yunshancheng@physics.ucla.edu}
\affiliation{Department of Physics and Astronomy, University of
  California, Los Angeles, California 90095, USA}
\author{Gang Wang}
\affiliation{Department of Physics and Astronomy, University of
  California, Los Angeles, California 90095, USA}
  \author{Zhiming Li}
\affiliation{Key Laboratory of Quark and Lepton Physics (MOE) and
Institute of Particle Physics, Central China Normal University, Wuhan 430079, China}
\author{Yuanfang Wu}
\affiliation{Key Laboratory of Quark and Lepton Physics (MOE) and
Institute of Particle Physics, Central China Normal University, Wuhan 430079, China}

\begin{abstract}
Hyper-order cumulants $C_5/C_1$ and $C_6/C_2$ of net-baryon distributions are anticipated to offer crucial insights into the phase transition from quark-gluon plasma to hadronic matter in heavy-ion collisions.
However, the accuracy of $C_5$ and $C_6$ is highly contingent on the fine shape of the distribution's tail, the detectable range of which could be essentially truncated by low statistics. 
In this paper, we use the fast Skellam-based simulations, as well as the Ultrarelativistic Quantum Molecular Dynamics model, to assess the impact of limited statistics on the measurements of $C_5/C_1$ and $C_6/C_2$ of net-proton distributions at lower RHIC energies.
Both ratios decrease from the unity baseline as we reduce statistics, and could even turn negative without a pertinent physics mechanism.
By incorporating statistics akin to experimental data, we can replicate the net-proton $C_5/C_1$ and $C_6/C_2$ values
comparable to the corresponding measurements for Au+Au collisions at $\sqrt{s_{NN}} =$ 7.7, 11.5 and 14.5 GeV.
Our findings underscore a caveat to the interpretation of the observed beam energy dependence of hyper-order cumulants.
\end{abstract}


\maketitle
\section{Introduction}
Cumulants, especially those of higher order, for conserved quantities in relativistic heavy-ion collisions, are crucial observables to probe the phase structure of Quantum Chromodynamics (QCD)~\cite{cumulant-1}. 
Conventionally, the cumulants are expressed as
\begin{eqnarray}\label{Cx-definition}
C_1 &=& \langle N \rangle, \\
C_2 &=& \langle (\delta N)^2 \rangle,  \\
C_3 &=& \langle (\delta N)^3 \rangle,  \\
C_4 &=& \langle (\delta N)^4 \rangle -3\langle (\delta N)^2 \rangle^2, \\ 
C_5 &=& \langle (\delta N)^5 \rangle - 10  \langle (\delta N)^2 \rangle \langle (\delta N)^3 \rangle, \label{eq:5}\\
C_6 &=& \langle (\delta N)^6 \rangle + 30 \langle (\delta N)^2 \rangle^3 \nonumber \\
& &-15\langle (\delta N)^2 \rangle \langle (\delta N)^4 \rangle -10 \langle (\delta N)^3 \rangle^2, \label{eq:6}
\end{eqnarray}
where $\delta N = N - \langle N \rangle$, $N$ is the number of particles in one event, and the average is taken over all events. In practice, cumulant ratios such as $C_2/C_1$, $C_3/C_2$, $C_4/C_2$, $C_5/C_1$, and $C_6/C_2$ are employed to mitigate the trivial volume dependence~\cite{ratio}. 
A Skellam distribution, defined as the difference between two independent Poisson distributions, is typically utilized as a statistical baseline~\cite{Skellam-2, Skellam-3}, which is unity for $C_4/C_2$, $C_5/C_1$, and $C_6/C_2$.

Net-baryon $C_5/C_1$ and $C_6/C_2$ are of great interest as
lattice QCD calculations~\cite{C6-karsch-v2} show that both of them are negative at low baryon chemical potential ($\mu_B$) near a pseudo-critical temperature, and
become more negative as $\mu_B$ increases. Additionally, the QCD-assisted low-energy effective theory and the QCD-based models, such as the Polyakov loop extended quark-meson model and the Nambu-Jona-Lasinio model, indicate that these two ratios are also sensitive to the chiral crossover transition~\cite{PQM-C6-v0,PQM-C6-v1,PQM-C6-v2, PNJL-C6-v1, PNJL-C6-v2}. 
In experiments, neutrons are usually undetectable, and net-proton number ($\Delta N_p$) is measured as a proxy for net-baryon number~\cite{proxy}.
The STAR Collaboration has reported the beam energy and collision centrality dependence of net-proton $C_5/C_1$ and $C_6/C_2$ in Au+Au collisions at center-of-mass energies ($\sqrt{s_{NN}}$) from 3 to 200 GeV~\cite{STAR-C6C2-v3, STAR-C6C2-v4}. Except at 3 GeV, the measured $C_6/C_2$ values for 0--40\% centrality collisions display a progressively negative trend with decreasing energy, resembling the aforementioned lattice QCD calculations. Conversely,
$C_5/C_1$ for 0--40\% centrality collisions from 7.7 to 200 GeV exhibits a weak beam energy dependence and fluctuates about zero.

Before connecting the measured $C_5/C_1$ and $C_6/C_2$ to lattice QCD calculations, however, it is vital to eliminate contributions unrelated to the phase transition. For instance, 
the hadron resonance gas model can also produce negative net-baryon $C_6/C_2$ values at $\sqrt{s_{NN}} \lesssim$ 40 GeV by integrating global baryon-number conservation~\cite{pbm-npa-conservation}.
Similar negative behaviors in $C_6/C_2$ are also noted in the sub-ensemble acceptance method~\cite{sam-1, sam-2, sam-3} and hydrodynamics calculations~\cite{hydrodynamics}, encompassing global conservation, system non-uniformity, and momentum space acceptance. 
Moreover, it is imperative to consider other mechanisms, including initial volume fluctuations~\cite{C6C2-volume-1, C6C2-volume-2, C6C2-volume-3}, as well as distinctions between net-proton and net-baryon cumulants~\cite{koch-1, asakawa-1, asakawa-2}.

Beyond physics mechanisms,
insufficient statistics can also affect the hyper-order
cumulant measurements~\cite{chenlz-c6-2021,chenlz-statistics-1, chenlz-statistics-2, chenlz-statistics-3}.
Cumulants reflect the shape of the distribution.
For example, skewness reveals left-right asymmetry, and kurtosis quantifies the peakedness of the distribution or the flatness of the tail.  
Hyper-skewness $\left(C_5/C_1\right)$ and hyper-kurtosis $\left(C_6/C_2\right)$ are particularly sensitive to the fine details of the distribution's tail. When the data analyses involve low statistics, the distribution's tail may not be fully explored and can be essentially truncated. The resultant hyper-order cumulants are thus distorted.
In this scenario, the observed $C_5/C_1$ and $C_6/C_2$ can be negative even with statistical Skellam-based simulations.
Additionally, to suppress initial volume fluctuations, data analyses commonly apply the centrality bin width correction (CBWC)~\cite{cbwc-v3, cbwc-star,cbwc-v2}, by conducting statistics within each multiplicity bin before aggregating the results for a finite centrality interval.
This exacerbates the impact of low statistics on the measured hyper-order cumulants.

In Sec.~\ref{sec2}, we use a Skellam distribution to illustrate how the sampling of the distribution's tail, or rather the lack thereof, influences the measured $C_5/C_1$ and $C_6/C_2$.  
In Sec.~\ref{sec3}, we corroborate the impact of limited statistics, especially the CBWC-induced effect, on both hyper-order cumulants using a more realistic model,  Ultrarelativistic Quantum Molecular
Dynamics (UrQMD)~\cite{UrQMD-1, UrQMD-2}.
In Sec.~\ref{sec4}, we present comprehensive Skellam-based Monte Carlo simulations with statistics akin to experimental data
at lower RHIC energies and 
compare the outcomes with the data. Finally, we provide a summary in Sec.~\ref{sec5}.

\section{Statistical Skellam Simulations}
\label{sec2}

We start the investigation of low-statistics effects with the statistical baseline using a
Skellam distribution,
\begin{equation}\label{Skellam-formula}
f(k,  \mu_1,  \mu_2) =  e^{- \left(\mu_1+\mu_2\right)} \left(\frac{\mu_1}{\mu_2}\right)^{\frac{k} {2} } I_{|k|} \left(2 \sqrt{\mu_1 \mu_2}\right),
\end{equation}
where $I_{|k|} (z)$ is the modified Bessel function of the first kind.
The input parameters,  $\mu_1$ = 31.217 and $\mu_2 = 0.96272$,
match the STAR measurements of $\left<N_{p}\right>$ and $\left<N_{\bar{p}}\right>$, respectively, in 0--5\% most central Au+Au collisions at $\sqrt{s_{NN}}$ = 11.5 GeV~\cite{cbwc-v3}, in which the number of analyzed events is around $10^6$. 

\begin{figure}[bhtp]
\centering
\includegraphics[width=0.48\textwidth]{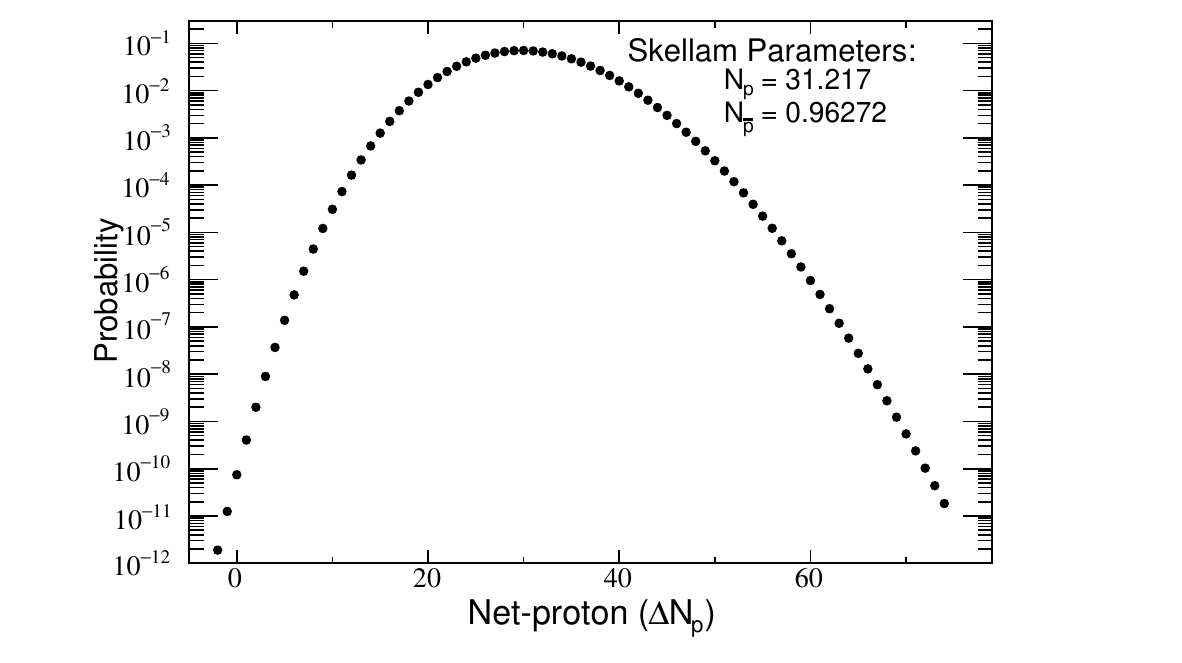}
\caption{\label{Skellam} Probability of net-proton number ($\Delta N_p$) based the Skellam distribution in Eq.~(\ref{Skellam-formula}). }
\end{figure}

\begin{figure}[bhtp]
\centering
\includegraphics[width=0.49\textwidth]{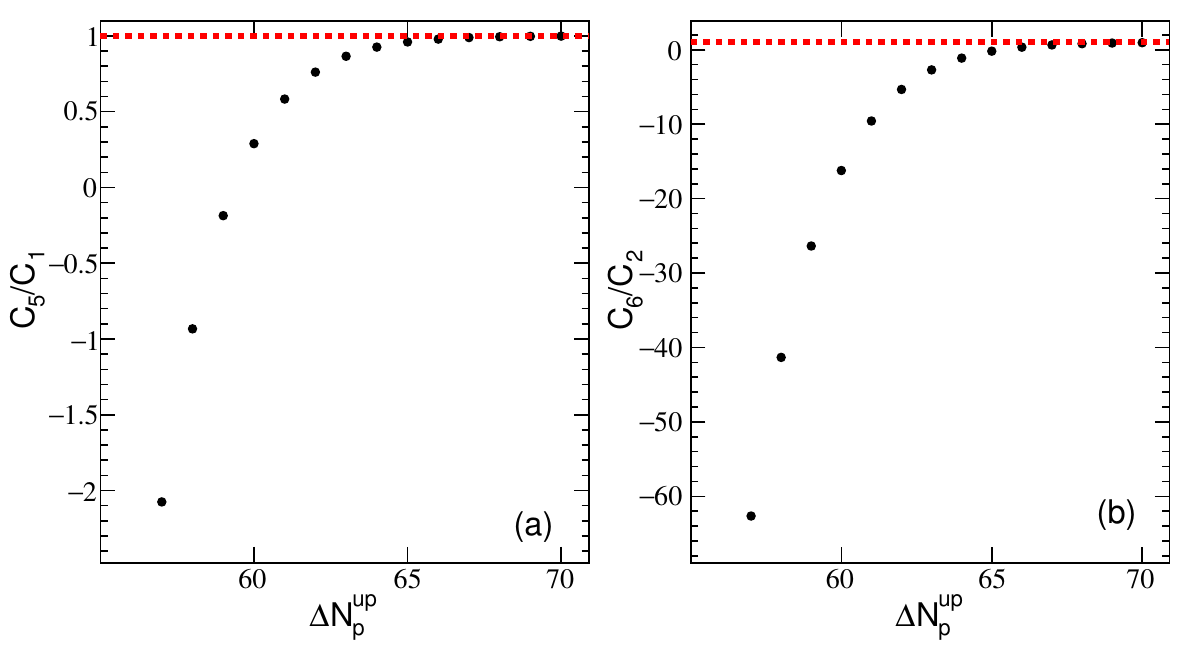}
\caption{\label{C6-tail-effect}  
(a) $C_5/C_1$ and (b) $C_6/C_2$ calculated with artificial truncation on the distribution in Fig.~\ref{Skellam}, excluding
events with $\Delta N_p$ exceeding the upper bound, $\Delta N_p^{\rm up}$. The red dashed lines at unity serve as the expected baselines.}
\end{figure}

Figure~\ref{Skellam} displays the normalized $\Delta N_p$ distribution according to  Eq.~(\ref{Skellam-formula}).    
With total statistics of $10^6$ events, detecting an event with a probability below $10^{-7} \sim 10^{-6}$ is rather improbable.
The region with $\Delta N_p \ge$ 58 has a cumulative probability of roughly $6.35 \times 10^{-6}$, suggesting an expectation of only 6--7 events occurring within this tail. This tail has a negligible impact on lower-order cumulants, but not on hyper-order ones.
For instance, its contribution to $C_1$ ($\approx$ 30.2543) is only 0.0004, constituting a relative  0.0013\%. 
However, 
an extra event with $\Delta N_p = 60$ will already add 24.3 and 729 to the first terms of $C_5$ ($= 
C_1 \approx 30.2543$) and $C_6$ ($\approx 32.1792$), respectively, as defined in Eqs.~(\ref{eq:5}) and (\ref{eq:6}).
Let alone other terms of $C_5$ and $C_6$.
In the left tail, events with $\Delta N_p$ between -2 and 6 have probabilities ranging from $10^{-12}$ to $10^{-7}$, and their contributions to $C_5$ and $C_6$ are minimal compared with those in the right tail. For the demonstration purpose, we will focus on events with $\Delta N_p$ in the right tail with probabilities below $10^{-6}$. 

Figures~\ref{C6-tail-effect}(a) and~\ref{C6-tail-effect}(b) show the truncation effect on ${C_5}/{C_1}$ and ${C_6}/{C_2}$, respectively, which are calculated by excluding events with $\Delta N_p$ exceeding a certain upper bound, $\Delta N_p^{\rm up}$. When $\Delta N_p^{\rm up}$ is around 70, both ${C_5}/{C_1}$ and ${C_6}/{C_2}$ 
adhere closely to the unity baseline, indicated by red dashed lines. As we reduce $\Delta N_p^{\rm up}$, both hyper-order cumulants decrease and could even turn negative with severe truncation. With $\Delta N_p^{\rm up} =$ 60, the resulting ${C_6}/{C_2}$ surpasses in magnitude the aforementioned physics mechanisms unrelated to the phase transition. 

In experiments, events in the right tail might be missing from analyses due to detector limitations or event selection. To examine this effect, we randomly generate 30 independent subsamples, each with $10^6$ events following the same Skellam distribution, as depicted in Fig.~\ref{Skellam}. From each subsample, we randomly reject one event with $\Delta N_p > 58$. Figure~\ref{C6-tail-truncate} presents the $C_6/C_2$ values before (solid circles) and after (open circles) removing one event.
The latter values
are consistently lower than the former.
This effect is expected to be more pronounced for even higher-order cumulants, such as $C_8$, necessitating careful treatments in data analyses.

\begin{figure}
\centering
\includegraphics[width=0.40\textwidth]{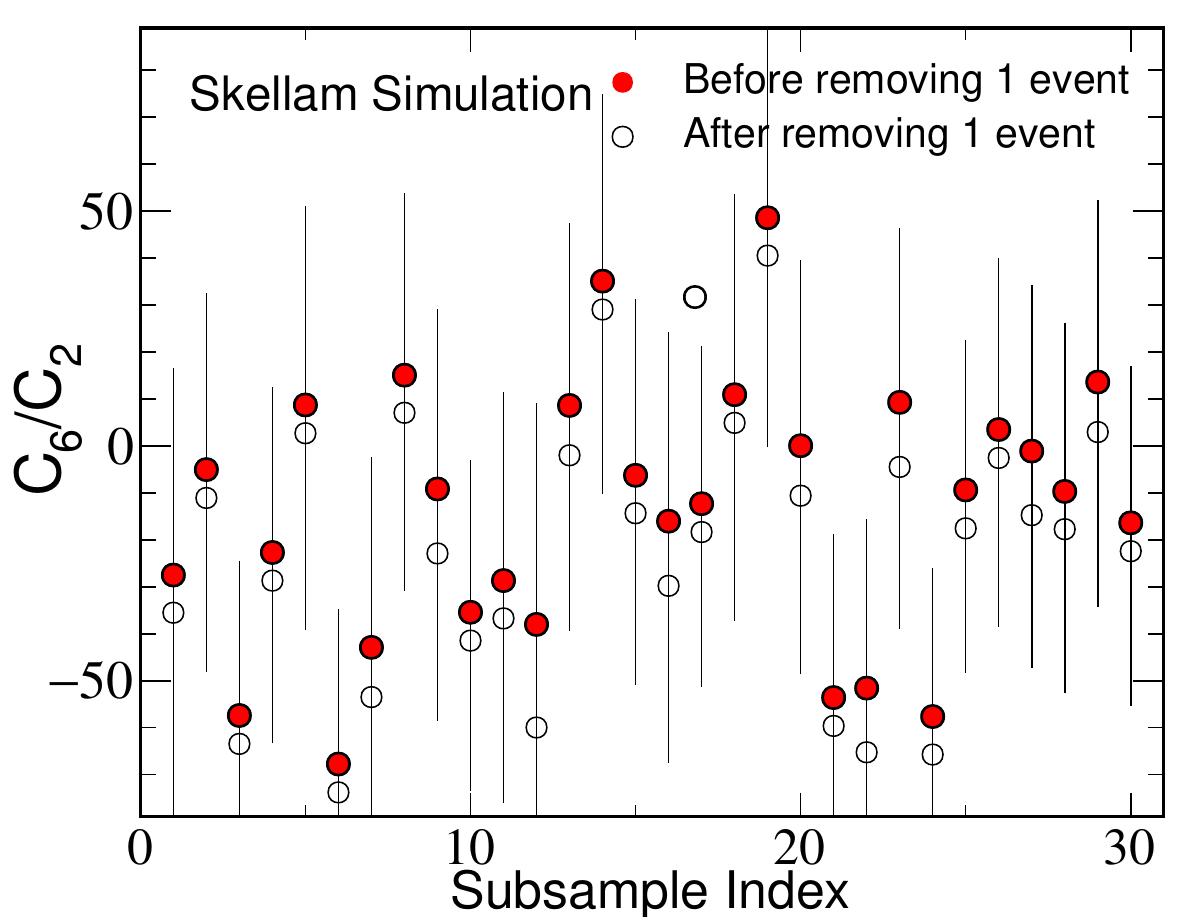}
\caption{\label{C6-tail-truncate} Skellam-based Monte Carlo simulations of $C_6/C_2$  (a) before and (b) after randomly rejecting one event with $\Delta N_p > 58$ from each subsample of $10^6$ events.}
\end{figure}

\section{UrQMD simulations for Au+Au Collisions at $\sqrt{s_{NN}}$ = 11.5 GeV}
\label{sec3}

The CBWC method in data analyses divides events from each centrality interval into finer multiplicity bins, leading to an essential reduction in the effective statistics and thus exacerbating the low-statistics effect on the measurements of hyper-order cumulants. We examine the CBWC-induced low-statistics effect on net-proton $C_5/C_1$ and $C_6/C_2$ using the UrQMD model (version 3.4)~\cite{UrQMD-1, UrQMD-2}.
In this study, multiplicity is quantified with the so-called RefMult3, the number of charged pions and kaons within the pseudorapidity range of $|\eta| < 1$ in each event, to avoid auto-correlations~\cite{centrality-definition}.
For a given centrality, cumulants ($C_1$, $C_2$, $C_5$, and $C_6$) are initially computed within each RefMult3 bin and then averaged over all RefMult3 bins, with the event number serving as a weight.
The kinematic regions for (anti)protons are $0.4<p_{T}<2$ GeV/$c$ and $|y|<0.5$. The statistical errors are estimated using the Bootstrap method~\cite{Bootstrap-1, Bootstrap-2}.

\begin{figure}[bhtp]
\centering
\includegraphics[width=0.48\textwidth]{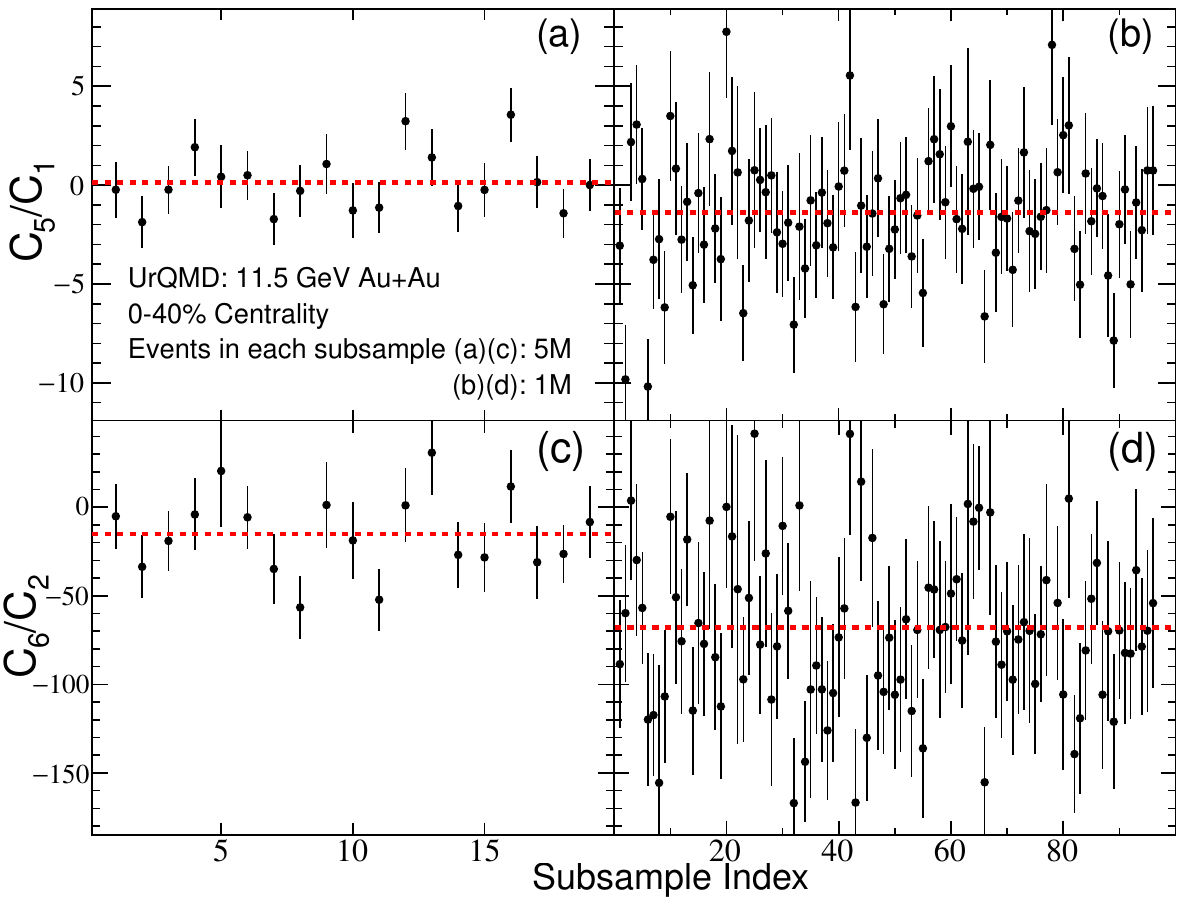}
\caption{\label{c62_compare_1m_5m_method1}UrQMD calculations of net-proton  $C_5/C_1$ (upper panels) and  $C_6/C_2$ (lower panels) in 0--40\% Au+Au collisions at $\sqrt{s_{NN}} =$ 11.5 GeV.
The entire sample of $9.6\times 10^7$ events is divided into subsamples of $5\times 10^6$ (left panels) or $10^6$ (right panels) events. The red dashed lines represent the ensemble averages over all subsamples.}
\end{figure}

We have generated $9.6\times 10^7$ UrQMD events for 0--40\% Au+Au collisions at $\sqrt{s_{NN}}$ = 11.5 GeV, matching the same statistics as experimental data~\cite{STAR-C6C2-v4}.
The complete sample is randomly divided into 19 or 96 subsamples, each containing $5\times10^6$ (5M) or $10^6$ (1M) events.  Figures~\ref{c62_compare_1m_5m_method1}(a) and~\ref{c62_compare_1m_5m_method1}(b) depict the net-proton $C_5/C_1$ values from the 5M and 1M subsamples, respectively, and Figs.~\ref{c62_compare_1m_5m_method1}(c) and~\ref{c62_compare_1m_5m_method1}(d) delineate the corresponding results for $C_6/C_2$. 
The red dashed lines represent the ensemble averages in each scenario, denoted by $\left<C_5/C_1\right>_{5{\rm M}}$, $\left<C_5/C_1\right>_{1{\rm M}}$, $\left<C_6/C_2\right>_{5{\rm M}}$, and $\left<C_6/C_2\right>_{1{\rm M}}$, respectively. 
Both $\left<C_6/C_2\right>_{5{\rm M}}$ and $\left<C_6/C_2\right>_{1{\rm M}}$ are negative in UrQMD simulations, without invoking any phase transition. 
Moreover, $\left<C_6/C_2\right>_{1{\rm M}}$ is significantly lower than $\left<C_6/C_2\right>_{5{\rm M}}$, corroborating the CBWC-induced low-statistics effect.
The CBWC-related effect on $C_5/C_1$ is also noteworthy,  albeit to a lesser extent. 
It is reasonable to assume that higher-order cumulants are more vulnerable to the low-statistics effect.

\section{ Statistics dependence of $C_5/C_1$ and $C_6/C_2$ at Lower RHIC energies}
\label{sec4}

As UrQMD calculations are computationally intensive, we instead perform fast Skellam-based Monte Carlo simulations to comprehensively study the statistics dependence of net-proton $C_5/C_1$ and $C_6/C_2$.
We generate numerous subsamples, each comprising $5\times10^6$ (5M) or $10^6$ (1M) events. In each event,  RefMult3 follows the same distribution as from UrQMD, and $N_p$ and $N_{\bar{p}}$ obey Poisson distributions, with $\left<N_{p}\right>$ and $\left<N_{\bar{p}}\right>$, respectively, derived from the corresponding RefMult3 bin in UrQMD. 
In the calculation of net-proton hyper-order cumulants, the CBWC approach is applied based on RefMult3.

\begin{figure}
\centering
\includegraphics[width=0.50\textwidth]{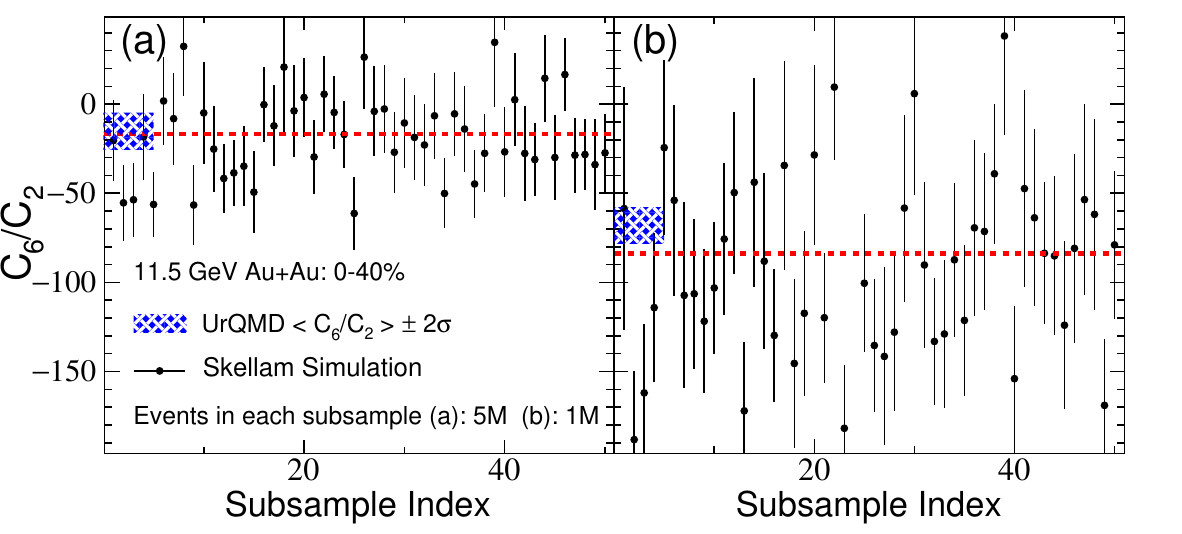}
\caption{\label{C62-UrQMD-simulations-5M} Skellam-based simulations of
net-proton $C_6/C_2$ for 0--40\% Au+Au collisions at $\sqrt{s_{NN}} = 11.5 $ GeV using subsamples of (a) $5\times10^6$ or (b) $10^6$  events. The  RefMult3 distribution as well as $\left<N_{p}\right>$ and $\left<N_{\bar{p}}\right>$ for each RefMult3 are taken from the UrQMD model. The red dashed lines represent the ensemble averages over 1000 subsamples.
The blue shaded bands denote the UrQMD calculations of $\left<C_6/C_2\right> \pm 2\sigma$ extracted from Figs.~\ref{c62_compare_1m_5m_method1}(a) and \ref{c62_compare_1m_5m_method1}(b).
  }
\end{figure}

Figure~\ref{C62-UrQMD-simulations-5M} illustrates the Skellam-based Monte Carlo simulations of net-proton $C_6/C_2$  from 50 subsamples for both (a) 5M and (b) 1M scenarios
in 0--40\% Au+Au collisions at 11.5 GeV.
The red dashed lines represent $\left<C_6/C_2\right>_{5{\rm M}}$ and $\left<C_6/C_2\right>_{1{\rm M}}$,  each averaged over 1000 subsamples. Both $\left<C_6/C_2\right>_{5{\rm M}}$ and $\left<C_6/C_2\right>_{1{\rm M}}$ are significantly below zero, while $\left<C_6/C_2\right>_{5{\rm M}}$ is higher than $\left<C_6/C_2\right>_{1{\rm M}}$.
For comparison, we also show the UrQMD results of $\left<C_6/C_2\right> \pm 2\sigma$ extracted from Figs.~\ref{c62_compare_1m_5m_method1}(a) and \ref{c62_compare_1m_5m_method1}(b), indicated with blue shaded bands.
Despite differences in event-generating mechanisms, the net-proton $\left<C_6/C_2\right>$ results exhibit remarkable similarities between UrQMD and Skellam-based simulations, both of which have the same statistics and the same parametrization for RefMult3, $\left<N_{p}\right>$, and $\left<N_{\bar{p}}\right>$.
This implies that the $C_6/C_2$ behavior in UrQMD is dominated by inadequate statistics. 
In other words, the details in the particle production mechanism cannot be properly revealed by net-proton $C_6/C_2$ given the currently available statistics.

\begin{figure}
\centering
\includegraphics[width=0.49\textwidth]{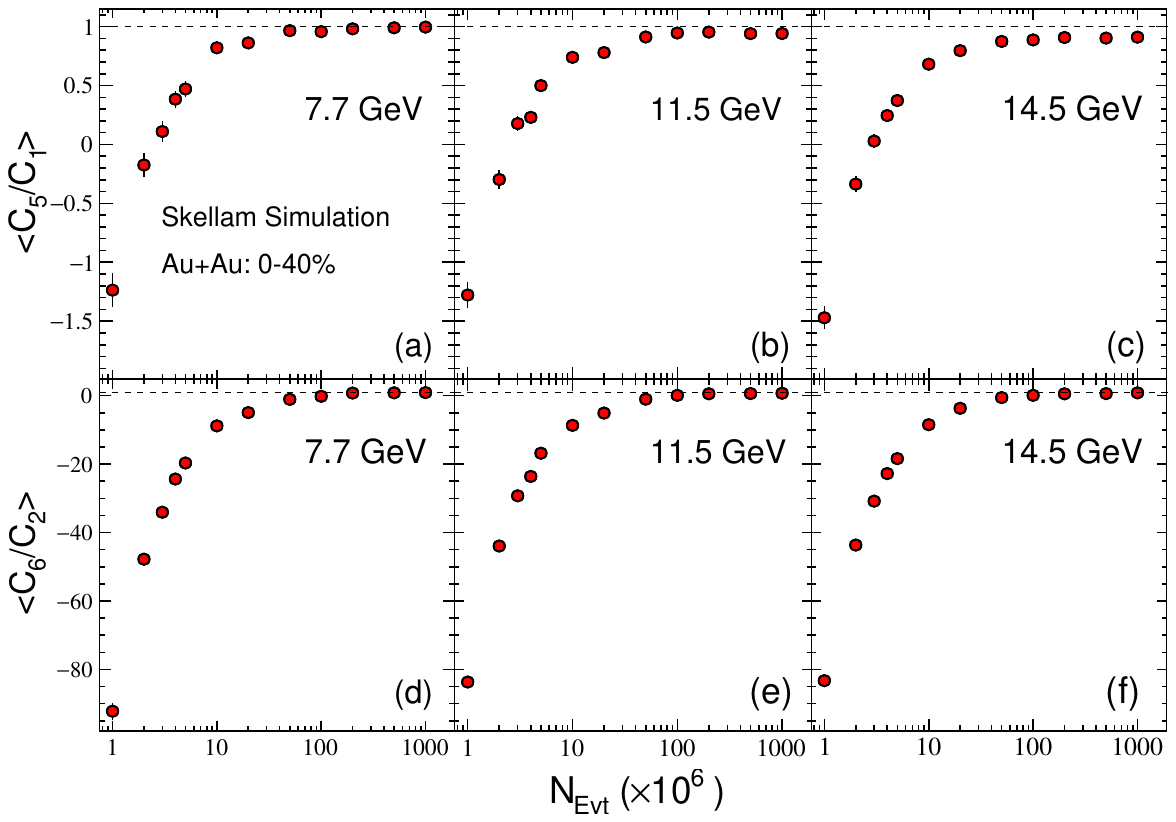}
\caption{\label{C62-UrQMD-simulations}  Statistics dependence of net-proton $\left<C_5/C_1\right>$ (upper panels) and $\left<C_6/C_2\right>$ (lower panels) in 0--40\% Au+Au collisions at $\sqrt{s_{NN}} = 7.7, 11.5 $ and 14.5 GeV from Skellam-based Monte Carlo simulations. $N_{\rm Evt}$ denotes the number of events in each subsample. $\left<C_5/C_1\right>$ and $\left<C_6/C_2\right>$ are computed by averaging over 1000 subsamples. The dashed lines at unity serve as the expected baselines.}
\end{figure}

As the Skellam-based simulation captures the essence of the low-statistics effect on hyper-order cumulant measurements, we employ it to explore the statistics dependence of net-proton $C_5/C_1$ and $C_6/C_2$ for 0--40\% Au+Au collisions at $\sqrt{s_{NN}}$ = 7.7, 11.5, and 14.5 GeV. The $C_5/C_1$ and $C_6/C_2$ results are presented in Figs.~\ref{C62-UrQMD-simulations}(a)--(c) and Figs.~\ref{C62-UrQMD-simulations}(d)--(e), respectively.
$N_{\rm Evt}$ denotes the number of events in each subsample, and $\left<C_5/C_1\right>$ and $\left<C_6/C_2\right>$ are computed by averaging over 1000 subsamples.
At all energies, $\left<C_6/C_2\right>$ increases with the growth of statistics as expected, and
analyses with $N_{\rm Evt} < 2\times10^7$ render significantly negative $\left<C_6/C_2\right>$ values. For the same statistics, the $\left<C_6/C_2\right>$ values at different beam energies are close to each other because the input parameters from UrQMD are akin among these energies.
$\left<C_6/C_2\right>$ approaches zero at $N_{\rm Evt} = 10^8$ and then turns positive but remains below unity at $N_{\rm Evt} = 2\times 10^8$. This can be attributed to the CBWC procedure, which divides events from each centrality interval into finer RefMult3 bins and reduces the effective statistics in the analysis. 
Compared with $\left<C_6/C_2\right>$,  $\left<C_5/C_1\right>$ deviates from unity to a lesser extent. At all three beam energies, the $\left<C_5/C_1\right>$ values are negative at $N_{\rm Evt} < 3\times10^6$, and approach unity at $N_{\rm Evt} > 5\times10^7$. 

\begin{figure}
\centering
\includegraphics[width=0.48\textwidth]{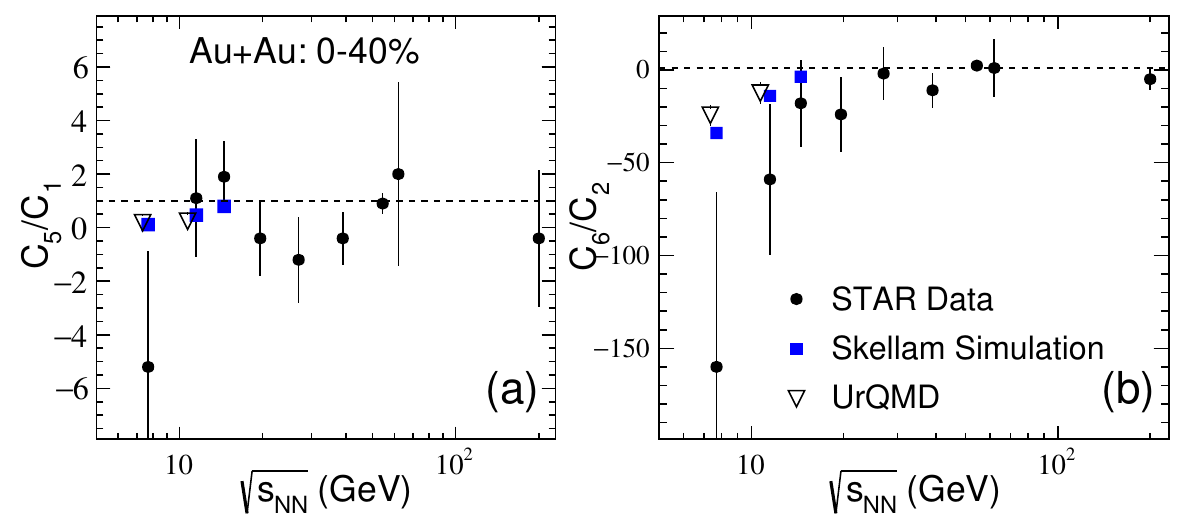}
\caption{\label{C51-C62-experiment} Skellam-based simulations of the beam energy dependence of net-proton (a) $\left<C_5/C_1\right>$ and   (b) $\left<C_6/C_2\right>$ in 0--40\% Au+Au collisions.
For comparison, we also show the STAR measurements~\cite{STAR-C6C2-v3, STAR-C6C2-v4} that have the same statistics as the simulations at the overlapping energies. The UrQMD calculations for $\sqrt{s_{NN}} =$ 7.7 and 11.5 GeV are slightly offset horizontally to improve clarity. The dashed lines at unity serve as the statistical baselines.}
\end{figure}

The numbers of events involved in the measurements of net-proton hyper-order cumulants during the RHIC Beam Energy Scan phase I (BES-I) are approximately $3\times10^6$, $6.6\times10^6$, and $2\times10^7$ in 0--40\% Au+Au collisions at $\sqrt{s_{NN}}$ = 7.7, 11.5 and 14.5 GeV, respectively~\cite{STAR-C6C2-v4}. 
Based on the discussions related to Fig.~\ref{C62-UrQMD-simulations}, it is evident that these datasets are insufficient to yield reliable measurements for $C_5/C_1$ and $C_6/C_2$.
Figure~\ref{C51-C62-experiment} shows the beam energy dependence of net-proton (a) $\left<C_5/C_1\right>$ and   (b) $\left<C_6/C_2\right>$ in 0--40\% Au+Au collisions from the Skellam-based simulations (blue solid squares).
The results are averaged over 1000 subsamples, each having the same statistics as the experimental data at the corresponding energy.
The UrQMD calculations (black open triangles) are plotted for 0--40\% Au+Au at 7.7(11.5) GeV, and the results are averaged over 35(14) subsamples, each containing $3\times10^6$($6.6\times10^6$) events. The STAR BES-I data (black solid circles)~\cite{STAR-C6C2-v4} are also presented for comparison, with the statistical and systematic uncertainties combined in quadrature.
The $C_5/C_1$ values from the Skellam-based simulations and UrQMD calculations are predominantly influenced by the low-statistics effect, ranging between 0 and 1, and aligning with the data within the associated uncertainties at all overlapping energies.
The $C_6/C_2$ values from both the Skellam-based simulations and UrQMD exhibit an ascending trend with beam energy and are
quantitatively consistent with experimental data at all collision energies under study.

\section{Summary}
\label{sec5}

Net-proton hyper-order cumulants potentially convey essential information on the phase transition in heavy-ion collisions, but the experimental measurements could be significantly affected by low statistics.
We have conducted statistical Skellam simulations to illustrate that insufficient statistics can limit the detectable range of the net-proton ($\Delta N_p$) distribution. Consequently, the derived values of $C_5/C_1$ and $C_6/C_2$ decrease from the unity baseline and could even turn negative without any discernible underlying physics.
In reality, even if the $\Delta N_p$ distribution is not severely truncated, 
a portion of the distribution's tail might be overlooked due to
detector inefficiency or event selection. 
Accordingly, we have further attempted to exclude one event with $\Delta N_p > 58$ from each subsample of one million events, resulting in significantly lower $C_6/C_2$ values than the original ones. 

In practice, to suppress initial volume fluctuations,
the centrality bin width correction (CBWC) necessitates statistical analyses within each multiplicity bin instead of a finite centrality interval, further amplifying the impact of low statistics. 
We have demonstrated the CBWC-related effect on net-proton $C_5/C_1$
and $C_6/C_2$ using UrQMD calculations for the 0--40\% centrality range in Au+Au collisions at 
$\sqrt{s_{NN}}$ = 11.5 GeV.
The corresponding influence on $C_6/C_2$ is notably stronger than on $C_5/C_1$.
The similarity between the results from UrQMD calculations and Skellam-based Monte Carlo simulations suggests that the hyper-order cumulants measured with the statistics involved in this analysis are dominated by the low-statistics effect.

We have further performed fast Skellam-based simulations with RHIC BES-I statistics for 0--40\% Au+Au collisions at $\sqrt{s_{NN}}$ = 7.7, 11.5, and 14.5 GeV.
The resultant net-proton $C_5/C_1$ spans from zero to unity, whereas the  $C_6/C_2$ values are significantly negative.
Both $C_5/C_1$ and $C_6/C_2$ are consistent with their corresponding experimental results within the uncertainties.
Specifically, the $C_6/C_2$ values in the simulations
distinctly exhibit a  decreasing trend as the collision energy decreases from
$\sqrt{s_{NN}} = 14.5$ to 7.7 GeV.
The concurrence between net-proton hyper-order cumulants from Monte Carlo simulations and real data, both in magnitudes and trends, raises a cautionary note in the interpretation of the observed beam energy dependence at RHIC. It is reasonable to speculate that even higher-order cumulants, such as $C_8$, will be more susceptible to the low-statistics effect.

The number of collision events collected at RHIC BES-II is typically 10--20 times higher than that at BES-I for each beam energy, and hence the low-statistics effect will be significantly reduced. According to the study presented in this paper, we anticipate that future measurements of both $C_5/C_1$ and $C_6/C_2$ from BES-II will show a notable increase compared with the current data.
To devise a method for mitigating the impact of limited statistics on the measured hyper-order cumulants, additional theoretical efforts are warranted to develop a realistic event generator that incorporates the physics of the phase transition, along with other mechanisms such as baryon conservation and proton-antiproton correlations.

\section{Acknowledgements}

This work is supported by the National Key Research and Development Program of China (No. 2022YFA1604900); the National Natural Science Foundation of China (No. 12275102), and the Open Fund of Key Laboratory of the Ministry of Education for the Central China Normal Universities (No. QLPL2021P01).
Y. C. and G. W. are supported
by the U.S. Department of Energy under Grant No. DE-FG02-88ER40424 and by the National Natural Science Foundation of China under Contract No.1835002.
We also acknowledge the High Performance Computing Center of Nanjing University of Information Science and Technology for their support of this work.

\ed